# Polarity-induced selective area epitaxy of GaN nanowires

Ziani de Souza Schiaber,[†,¶] Gabriele Calabrese,[‡] Xiang Kong,[‡] Achim Trampert,[‡] Bernd Jenichen,[‡] José Humberto Dias da Silva,[†] Lutz Geelhaar,[‡] Oliver Brandt,[‡] and Sergio Fernández-Garrido*,[‡]

[†]*Laboratório de Filmes Semicondutores Universidade Estadual Paulista Bauru, 17033-360 São Paulo, Brazil*
[‡]*Paul-Drude-Institut für Festkörperelektronik, Hausvogteiplatz 5–7, 10117 Berlin, Germany*
[¶]*On leave at Paul-Drude-Institut für Festkörperelektronik from her home institution*

E-mail: garrido@pdi-berlin.de

## Abstract

We present a conceptually novel approach to achieve selective area epitaxy of GaN nanowires. The approach is based on the fact that these nanostructures do not form in plasma-assisted molecular beam epitaxy on structurally and chemically uniform cation-polar substrates. By *in situ* depositing and nitridating Si on a Ga-polar GaN film, we locally reverse the polarity to induce the selective area epitaxy of N-polar GaN nanowires. We show that the nanowire number density can be controlled over several orders of magnitude by varying the amount of pre-deposited Si. Using this growth approach, we demonstrate the synthesis of single-crystalline and uncoalesced nanowires with diameters as small as 20 nm. The achievement of nanowire number densities low enough to prevent the shadowing of the nanowire sidewalls from the impinging fluxes paves the way for the realization of homogeneous core-shell heterostructures without the need of using *ex situ* pre-patterned substrates.

Single-crystalline GaN nanowires (NWs) can be synthesized by plasma-assisted molecular beam epitaxy (PA-MBE) using the self-assembled growth approach[1] on a wide variety of crystalline (e. g., AlN/Al$_2$O$_3$, Si, C, SiC, Ti, Mo), polycrystalline (Ti and Mo foils) and amorphous substrates (e. g., Si$_x$O$_y$, Al$_x$O$_y$).[2-11] In the absence of structural and morphological defects of the substrate, GaN NWs crystallize in the wurzite modification, exhibit well defined $M$-plane sidewall facets, and elongate along the [000$\bar{1}$] direction, namely, they are N-polar.[5,12-14] Although individual GaN NWs can be as thin as 15–20 nm[15] and are free of threading dislocations as well as homogeneous strain,[16] well-developed NW ensembles are typically characterized by a non-negligible degree of coalescence.[17] As NWs merge, their effective diameter increases[17] and crystal defects are generated at the coalesced joints, giving rise to small angle grain boundaries and inhomogeneous strain.[18-22] Coalescence is mainly caused by bundling of NWs, a phenomenon driven by the gain of surface energy at the expense of the elastic energy of bending.[23] This phenomenon becomes energetically favorable once the NWs reach a certain critical length and is facilitated by their spatially random nucleation in very high densities (typical values are on the order of 0.5–5×10$^{10}$ cm$^{-2}$).[15,24,25] Since the final NW number density is mainly determined by the shadowing of the substrate from the impinging fluxes by already existing NWs,[23] coalescence is an intrinsic characteristic of the self-assembled growth of GaN NWs in PA-MBE on chemically and structurally uniform substrates.

In order to overcome the fundamental limitations of the self-assembled growth approach, selective area epitaxy (SAE) was developed by several groups. To achieve control on the size and positioning of NWs, dielectric (e. g., $Si_xN_y$, $Si_xO_y$) or metallic (e. g., Ti, Mo) masks are patterned using *ex situ* lithographic techniques on GaN or Si substrates.[26–33] On these patterned substrates, SAE is achieved due to the higher GaN nucleation probability inside the holes opened in the mask. Using this growth approach, the fabrication of ordered arrays of GaN NWs free of coalescence and with narrow diameter distributions has been demonstrated.[26] The SAE of GaN NWs on *ex situ* pre-patterned substrates has, however, also several drawbacks with respect to the self-assembled growth approach: (i) the NW diameters attainable by SAE are limited by the lithographic resolution and the subsequent processing steps required to pattern the substrate, (ii) sub-30 nm patterning requires advanced lithographic techniques as well as state-of-the-art and fully optimized processing technology, making this approach substantially more complex and costly compared to self-assembly, (iii) the various technological steps introduce additional constraints regarding the choice of the substrate, and (iv) both the growth equipment and the NWs produced on patterned substrates are at a risk of contamination due to process residues.

In this letter, we present a conceptually novel SAE approach suitable to control the NW number density and prevent their coalescence without using *ex situ* pre-patterned substrates. Our approach consists of taking advantage of the fact that GaN NWs do not form at all on uniform cation-polar substrates.[5] By depositing Si on a Ga-polar GaN layer, we are able to locally reverse the polarity and thus to induce the SAE of N-polar GaN NWs. We demonstrate that the NW number density can be tuned over more than two orders of magnitude by varying the amount of deposited Si. Hence, this growth approach allows us to gain control over the nanowire number density while preserving the small diameters and the high structural perfection of self-assembled GaN NWs.

Our attempt of using Si to locally reverse the polarity of Ga-polar GaN films is motivated by the theoretical work of Rosa and Neugebauer[34] who investigated, within the framework of density functional theory, the structural and electronic properties of Si layers adsorbed on GaN(0001) surfaces. Their results suggest that under N- and Si-rich conditions, the formation of Si–N bonds may reverse the polarity of Ga-polar GaN films. In a previous work of some of the present authors,[5] this mechanism was invoked to explain the formation of N-polar GaN NWs on Al-polar AlN films on 6H-SiC(0001) substrates. However, a direct experimental test of these theoretical results has not yet been undertaken.

We first investigate the influence of the total amount of deposited Si on the possible formation and distribution of GaN NWs by SEM. Figures 1(a)–1(h) present exemplary bird's eye scanning electron micrographs of samples prepared using several Si deposition times $t_{Si}$ from 0 to 60 min. The total GaN growth time upon the nitridation of the Si layer was 180 min for the samples shown in Figs. 1(a)–1(d) and 360 min for the one shown in Figs. 1(e)–1(h). The micrographs presented in Figs. 1(e)–1(h) correspond all to the same sample but were taken at different positions across the wafer to illustrate the influence of a continuous variation in the amount of deposited Si. Note that this amount varies continuously from one edge to the other of the wafer by approximately 20% due to the lack of substrate rotation. As can be seen in Fig 1(a), in the absence of Si deposition, we observe the formation of a GaN layer. This layer is rather smooth despite of the use of N-rich growth conditions. In accordance with Refs. 35 and 36, this smooth surface is attributed to the simultaneous partial decomposition of the growing layer at 790°C. This phenomenon reduces the effective N surface coverage and may thus facilitate the layer-by-layer growth of GaN films under nominally N-rich growth conditions. As shown in Fig 1(b), for a Si deposition time of 10 min, the surface becomes rough but GaN NWs do not form. However, for a value of $t_{Si}$ of 21.5 min and above, we observe the formation of GaN NWs [see Figs. 1(c)–1(h)]. The NWs intersperse a faceted layer and their number density increases with the amount of deposited Si. Figures 1(c), 1(e) and 1(f) demonstrate that, for an optimized amount of deposited Si, it is possible to obtain ensembles of completely uncoalesced GaN NWs. As shown in the cross-sectional scanning electron micrograph presented in the inset of Fig. 1(e), these uncoalesced NWs exhibit small diameters (well below 30 nm), smooth sidewalls, and flat top facets.

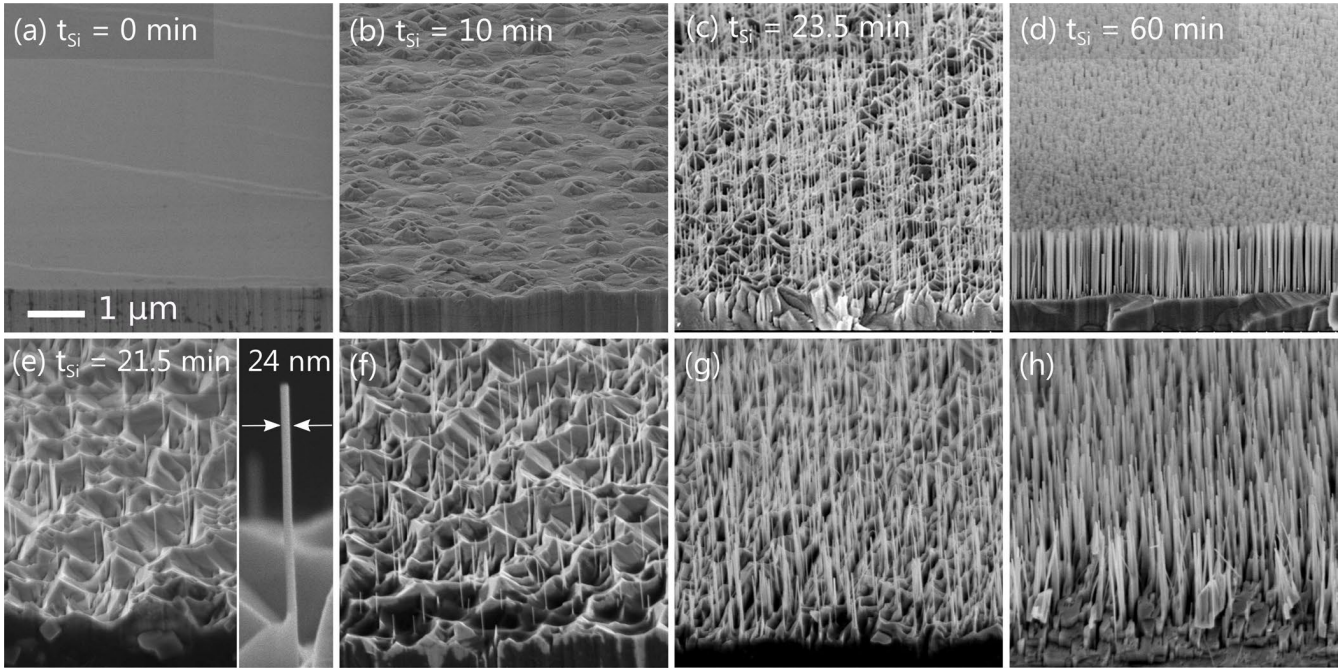

Figure 1: (a)–(d) Bird's eye view scanning electron micrographs of samples prepared using different Si pre-deposition times, $t_{Si}$, as indicated in the images. The GaN growth time upon Si nitridation for all these samples was 180 min. (e)–(h) Bird's eye view scanning electron micrographs illustrating the continuous variation of the morphology across the wafer for $t_{Si} = 21.5$ min. As result of the lack of substrate rotation, the effective amount of pre-deposited Si increases from (e) to (h). The inset in (e) presents a highly magnified cross-sectional view of a GaN NW. The GaN growth time for this sample was 360 min. All the images but the inset shown in (e) were acquired using the same magnification.

The quantitative variation of the NW number density with the Si deposition time, derived from a statistical analysis of scanning electron micrographs as described in the Methods section, is shown in Fig. 2(a). We do not distinguish here between single NWs and coalesced aggregates, hence, the values shown in the graph represent the total number density of objects. For values of $t_{Si}$ lower than 21.5 min, NWs do not form. With increasing Si deposition time, the NW number density increases continuously but steeply from $1.7 \times 10^7$ to $9.3 \times 10^9$ cm$^{-2}$. According to our calibration of the Si cell, the critical equivalent Si layer thickness required to trigger the formation of GaN NWs is about 1.1 ML. Finally, for Si deposition times longer than approximately 25 min, the NW number density eventually decreases as the result of coalescence of closely spaced NWs.

Figure 2(b) presents the average diameter of all rod-like objects detected in the scanning electron micrographs as a function of the Si deposition time. This quantity increases from 20 to 80 nm when $t_{Si}$ is varied between 21.5 and 60 min. The increase in the average NW diameter with the Si deposition time is attributed to the increasing degree of coalescence associated to the concomitant variation in the NW number density [see Fig. 2(a)]. In fact, as can be seen in Fig. 2(c), for $t_{Si} = 23.5$ min, the diameter distribution in the areas of the wafer as those shown in Fig. 1(c) is narrow 8 nm), symmetric, and well-described by a normal distribution, as expected for ensembles of uncoalesced NWs. In contrast, for longer Si deposition times [see Fig. 2(c) for $t_{Si} = 60$ min], the diameter distribution becomes broad, asymmetric, and is represented well not by a normal but a gamma distribution, a clear fingerprint of NW coalescence.[17] The results shown above demonstrate that it is possible to induce the formation of ensembles of thin NWs with different number densities by depositing Si on a Ga-polar GaN layer. However, they do not demonstrate that the growth of these NWs is actually induced by a local inversion of crystal polarity. To gain further insights into the underlying formation mechanisms, the structural properties of the samples were further investigated by x-ray diffraction (XRD), transmission electron microscopy (TEM), convergent beam electron diffraction (CBED), and wet chemical etching.

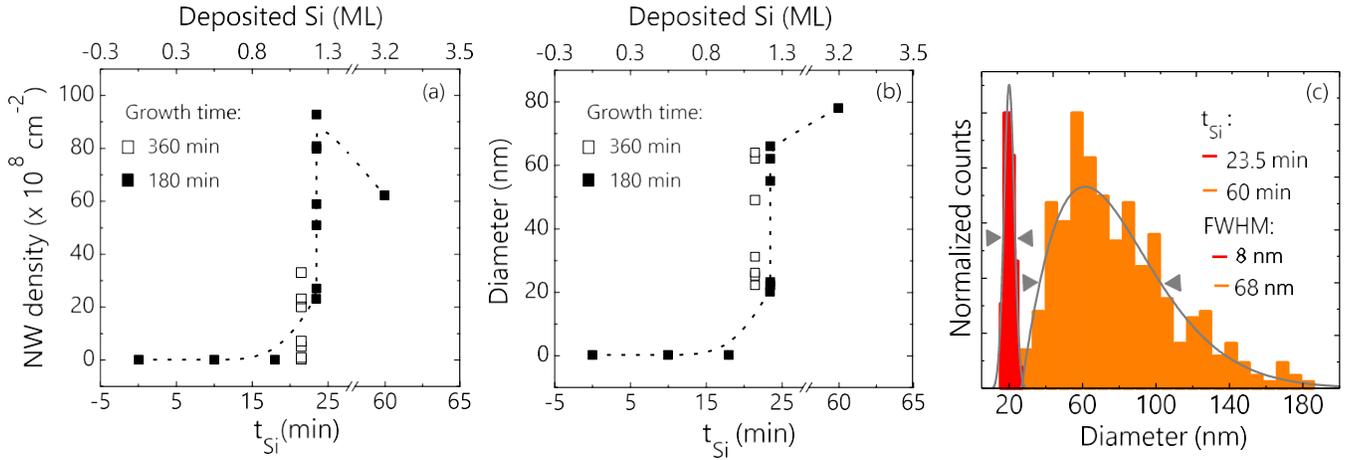

Figure 2: (a) Number density and (b) average diameter of all rod-like objects detected in scanning electron micrographs as a function of the amount of pre-deposited Si. For those samples exhibiting a non-negligible gradient of these quantities across the wafer, their different values were assigned to the corresponding nominal amounts of pre-deposited Si. The total GaN growth time upon Si nitridation for the different samples is indicated in the legends. The dashed lines in (a) and (b) are guides to the eye. (c) Diameter distribution of two samples prepared using different Si pre-deposition times, namely, 23.5 and 60 min. The bin sizes were selected according to the Freedman–Diaconis rule. The solid lines are fits of normal and shifted gamma distributions to the experimental data for $t_{Si}$ = 23.5 and 60 min, respectively. The full width at half maximum (FWHM) of the diameter distributions are indicated in the figure.

To analyze the epitaxial reationship between the GaN NWs and the GaN(0001) layer they emanate from, the sample prepared using a Si deposition time of 60 min [see Fig. 1(d)] was analyzed by XRD. If the GaN NWs are in registry with the GaN layer, the NW- and layer-related XRD reflections will coincide. However, if the orientation of the NWs differ from that of the layer, the high number density of the GaN NWs for this particular sample will facilitate their detection.[22] Figure 3(a) shows the symmetric θ/2θ scan performed to determine the out-of-plane orientation of the GaN NWs. Apart from the peaks related to the $Al_2O_3$(0001) substrate and the AlN(0001) buffer layer, we only observe the GaN 0002, 0003, 0004 and 0006 Bragg reflections. These reflections arise from the GaN NWs as well as the GaN layer. The lack of any additional GaN-related peak evidences that, as expected, the GaN NWs elongate along the (0001) axis. The in-plane orientation of the GaN NWs was elucidated by acquiring a pole figure of the GaN $10\bar{1}5$ reflections. If the GaN NWs nucleated on an amorphous Si or $Si_xN_y$ interlayer, they would exhibit a random in-plane orientation and we should then observe a ring intersecting the reflections from the GaN layer underneath. Nevertheless, as shown in Fig. 3(b), only six well-defined diffraction spots are detected, demonstrating the registry of the NWs with the GaN layer. Therefore, the GaN NWs do not nucleate on amorphous Si or $Si_xN_y$.

Figure 4(a) presents a representative transmission electron micrograph of the sample analyzed by XRD. Several GaN NWs are evident, but also the interface between the GaN template and the 100 nm thick GaN buffer layer grown by PA-MBE is clearly observed. Figures 4(b) and 4(c) show the experimental as well as simulated CBED pat-terns of the GaN buffer layer and an exemplary GaN NW, respectively. The CBED patterns confirm that the GaN layer is Ga polar and reveal that the GaN NWs grown on top are N polar. The polarity of the GaN NWs was further investigated on a macroscopic scale by KOH chemical etching. As shown in the inset of Fig. 4(a), most of the NWs were entirely dissolved upon exposing them to KOH, and the remaining ones exhibit a pencil-like shape. The high etch rate of the flat top facet of the NWs confirms their N polarity.[12,37] Hence, as intended and further discussed below, the polarity was reversed at the interface between the buffer layer and the GaN NWs due to Si deposition.

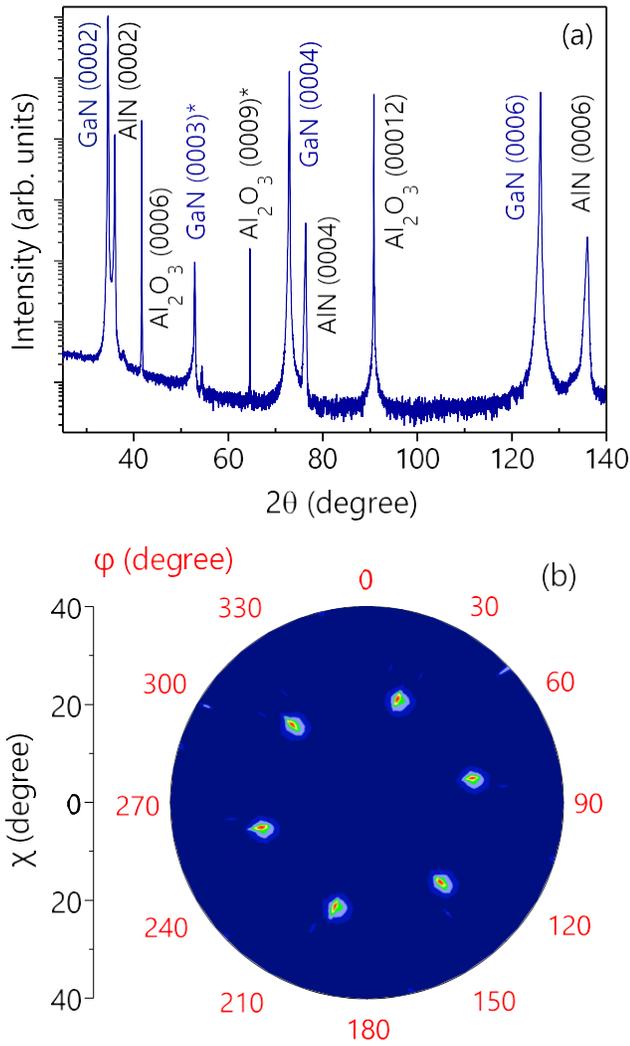

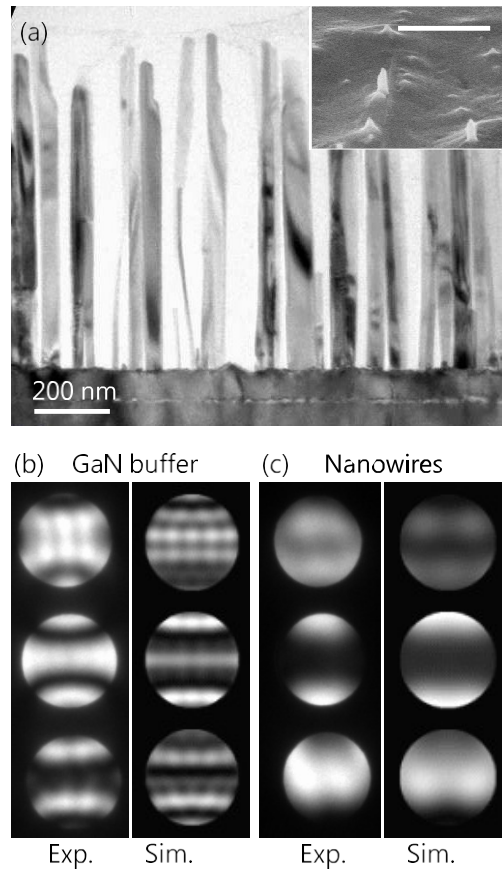

Figure 3: (a) Symmetric θ/2θ XRD scan and (b) XRD pole figure of the GaN $10\bar{1}5$ reflection for the sample prepared using a Si pre-deposition time of 60 min. The forbidden reflections in (a) are labeled with an asterisk.

Figure 4: (a) Cross-sectional transmission electron micrograph of the sample prepared using a Si pre-deposition time of 60 min. The inset shows a bird's eye view scanning electron micrograph acquired after exposing the sample to KOH. The scale bar in the inset represents 1 µm. Experimental and simulated CBED patterns of (b) the GaN buffer layer and (c) a GaN NW for the 0002, 0000, and $000\bar{2}$ disks (from bottom to top). The specimen thicknesses assumed for the simulations in (b) and (c) are 131 and 52 nm, respectively.

Figure 5(a) and 5(b) show highly-magnified bright-field transmission electron micrographs taken at the interface between the buffer layer and the GaN NWs of the sample previously investigated by XRD and TEM as shown in Figs. 3 and 4, respectively. As already seen in Fig. 1 and further discussed in the Methods and Supporting Information sections, the surface becomes rough and faceted as a result of the Si deposition at high temperature. It is clear from Figs. 5(a) that all GaN NWs emanate from a faceted layer. To elucidate the origin of the polarity inversion observed here [cf. Figs. 4(b) and 4(c)], we investigate the interface between the NWs and the faceted layer underneath by HRTEM. Figures 5(b) and 5(c) show HRTEM micrographs taken from the interface between two different NWs and the faceted layer (additional micrographs corresponding to other NWs can be found in the Supporting Information). In both cases, we clearly observe the presence of a 1–2 ML thick interlayer. Depending on the location, this interlayer is found to be disordered and close to amorphous or perfectly crystalline, as seen in Figs. 5(b) and 5(c), respectively.

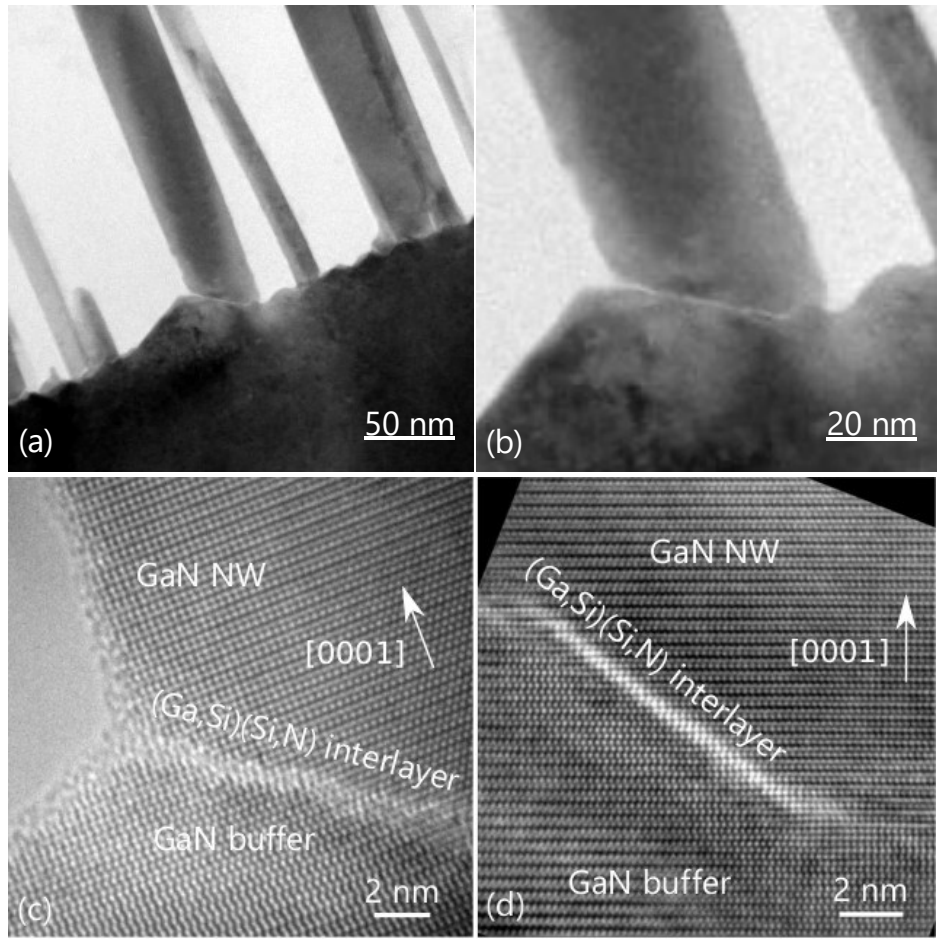

Figure 5: (a) and (b) Representative cross-sectional transmission electron micrograph of a GaN NW emerging from a facet of the GaN buffer layer. (c) and (d) High-resolution micrographs of the interface region for two different NWs recorded along the $(11\bar{2}0)$ zone axis. A 1–2 ML thick (Si,Ga)N interfacial layer is evident in both cases.

It is this interlayer that induces the inversion of crystal polarity and the subsequent formation of N-polar GaN NWs. Our results are reminiscent of the polarity inversion observed in GaN(0001) layers heavily doped with Mg.[38–40] In this case, for a sufficiently high Mg flux, the polarity is eventually reversed from Ga to N via the formation of a 1 ML thick faceted, coherent, and dislocation-free $Mg_3N_2$ interlayer. Obviously, in the present case, it would be also of great interest to identify the atomic structure and to determine the chemical composition of the interlayer. Concerning the latter point, we do not have direct access to its specific chemical composition, but the strong contrast variation across the interlayer observed in Fig. 5(c) implies a substantial concentration of substitutionally incorporated Si atoms giving rise to high local strain. By exposing the GaN(0001) surface to the Si flux at 790°C, we have thus locally formed a (Ga,Si)(Si,N) interlayer. The formation of such a compound can be understood by taking into account the amphoteric nature of Si in GaN as well as the fact that Si can also crystallize in a tetrahedrally bonded hexagonal modification.[41,42] Within this scenario, the inversion of crystal polarity might be caused by the insertion of a locally non-polar material which possesses the point- and space -group of Lonsdaleite. Regarding the former point, Fig. 6 shows that the interlayer forms a coherent boundary between the GaN buffer and the GaN NW. In particular, Fig. 6(a) demonstrates that the $(1\bar{1}00)$ lattice fringes are continuous across the boundary. Although local distortions are visible, no misfit dislocations are created at the interface. The same is true for the $(0002)$ lattice fringes as seen in Fig. 6(b). These fringes reflect the characteristic 2H stacking sequence of the wurtzite structure throughout the image, but closer inspection reveals them to exhibit a vertical shift when comparing the regions on the left and right of the boundary. In fact, the entire lattice is found to be shifted by $c/2$ at the boundary as indicated by the arrows. This shift is likely to be a direct consequence of the polarity inversion that occurs at this boundary. However, a full understanding of the mechanism of polarity inversion would require an in depth experimental and theoretical investigation of the atomistic nature of this particular inversion domain boundary. The important result in the present context is that we observe the formation of a coherent and crystalline (Ga,Si)(Si,N) interlayer that induces the polarity inversion and allows the formation of GaN NWs.

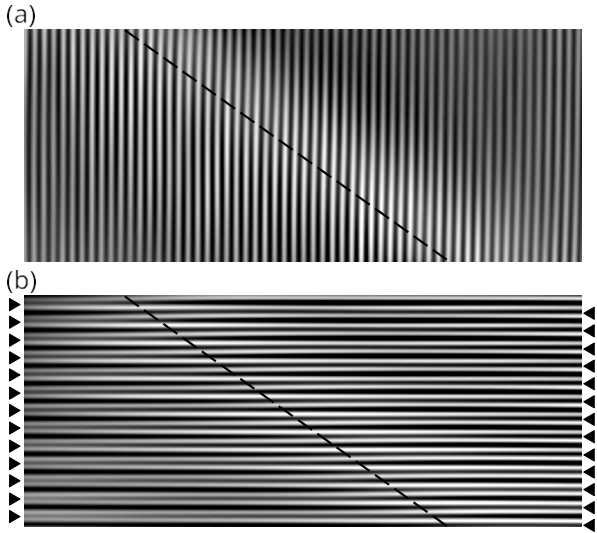

Figure 6: Magnified and Fourier-filtered section of the high-resolution micrograph depicted in Fig. 5(c) visualizing (a) the $(1\bar{1}00)$ and (b) the $(0002)$ lattice fringes. The diagonal dashed line indicates the position of the interface between the GaN buffer on the lower left and the GaN NW on the upper right. The arrows in (b) are separated by a distance equal to the $c$ lattice constant (i. e., 0.51855 nm) and highlight the shift of the lattice fringes at the interface by $c/2$.

To summarize and conclude, we have demonstrated a conceptually novel growth approach to achieve SAE of GaN NWs by PA-MBE without using *ex situ* pre-patterned substrates. In contrast to conventional SAE approaches, which rely on the higher GaN nucleation probability inside the holes opened in a mask, we here gain control over NW formation by taking advantage of the fact that these nanostructures do not form on structurally and chemically uniform cation-polar substrates. By *in situ* depositing Si on a Ga-polar GaN layer, we locally reverse its polarity to induce the formation of single-crystalline N-polar GaN NWs. Using this method, the NW number density can be tuned over several orders of magnitude by varying the amount of pre-deposited Si. The lowest NW number density that we have achieved is $1.7 \times 10^7$ cm$^{-2}$, i. e., more than two orders of magnitude lower than the typical values obtained for self-assembled GaN NWs. This dramatic reduction in NW density allowed us to fabricate ensembles of uncoalesced NWs with diameters as small as 20 nm. As reported in detail elsewhere,[43] further control on the NW diameter can be achieved by tuning the ratio between the impinging Ga and N fluxes. Interestingly, the lowest NW number densities reported in this work are sufficient to prevent the shadowing of the NW sidewalls from the impinging fluxes.[44] Therefore, our results paves the way for the realization of homogeneous core-shell NW heterostructures by PA-MBE without relying on *ex-situ* pre-patterned substrates as reported before in Refs. 45–47.

Obviously, the major drawback of the present growth approach is the lack of control over the positioning of GaN NWs. However, for those applications where the arrangement of GaN NWs is a matter of concern, the concept demonstrated here can be combined with the use of lithographic techniques to pre-define the positioning of GaN NWs. This combination, of course, comes at the expense of the inherent disadvantages associated to the use of pre-patterned substrates. For that purpose, a substrate compatible with the self-assembled for-mation of GaN NWs must be masked with a cation-polar layer or vice versa. The polarity-induced SAE of GaN NWs would then offer interesting fundamental advantages over previous SAE concepts: (i) regardless of growth time, NWs will never form on the cation-polar mask, (ii) due to the higher thermal stability of N-polar GaN, any kind of parasitic growth on the cation-polar mask can be eventually suppressed by using sufficiently high substrate temperatures[32,48] and, (iii) the patterned substrate does not require the presence of foreign elements.

Finally, it is worth noting that the findings reported in this manuscript may not only be relevant for the growth of group-III nitride NWs, but potentially also for the SAE of other polar semiconductor compounds. A prominent candidate is ZnO,[14,49] another material that, regardless of the substrate, exhibits a pronounced tendency to form NWs with a unique and well-defined polarity.

## Methods

### Growth

All samples were grown in a PA-MBE system equipped with solid source effusion cells for Ga and Si as well as a radio-frequency N plasma-source to generate active N. The impinging Ga and N fluxes were calibrated in units of ML/s as described in Ref. 50. A growth rate of 1 ML/s is equivalent to $1.14 \times 10^{15}$ atoms cm$^{-2}$ s$^{-1}$. The Si flux was also calibrated in ML/s by analyzing, using secondary ion mass spectrometry, the amount of Si incorporated into GaN films grown under intermediate metal-stable growth conditions[50] as a function of the Si cell temperature. As substrates, we used 5 μm thick Ga-polar GaN layers prepared by hydride vapor phase epitaxy on 2" Al$_2$O$_3$(0001) wafers with an AlN buffer layer (purchased from Kyma). The backside of the substrates was coated with an appoximately 1 μm thick Ti layer for efficient radiation coupling. Prior to the actual growth experiments, 2 ML of Ga were deposited at 690°C and flashed off to remove residual contaminants from the surface. Then, an approximately 100 nm thick GaN layer was grown under intermediate metal-stable growth conditions[50] to obtain a smooth and flat surface (see the corresponding atomic force micrograph in the supporting information) as well as to bury residual contamination on the substrate surface. Afterwards, the radio-frequency N plasma-source was switched off and the substrate temperature increased to 790°C. At this point, Si was deposited using a flux of $0.9 \times 10^{-3}$ ML/s obtained with a Si cell temperature of 1250°C. The Si deposition time was varied among the samples between 0 and 60 min. Hence, the corresponding equivalent nominal Si layer thicknesses were in the range of 0–3.2 ML. As shown in the Supporting Information, where we present atomic force micrographs of samples prior and after Si deposition, the deposition of Si results in the formation of three-dimensional islands. The density of the islands increases with the amount of deposited Si. Beside the formation of three-dimensional islands, the areas in between them become rougher than the original GaN layer. The overall roughing of the surface was also observed in-situ by reflection-high energy electron diffraction (RHEED) (see RHEED patterns in the Supporting Information). The originally streaky RHEED pattern evolves toward a spotty one during Si deposition. All the reflections observed in this pattern still originate from GaN, i. e., we do not observe the formation of cubic Si. It seems that the impinging Si atoms initiate a considerable mass-transport and diffuse into the GaN layer, a phenomenon that can be favored by the non-negligible GaN decomposition rate at 790°C.[48] Such a scenario is compatible with the transmission electron micrographs shown in Figs. 5(b), 5(b), and 6, where we detect the formation of a pseudomorphic (Ga,Si)(Si,N) alloy with the same crystal structure as the underneath GaN layer. After Si deposition, we ignited the radio-frequency N plasma-source and nitridated the surface for 2 min. Finally, we opened the Ga shutter to initiate the growth of GaN NWs. The Ga and N fluxes during the growth of the GaN NWs were 0.3 and 0.5 ML/s, respectively. During the growth experiments, we did not rotate the substrates.

## Characterization

The morphological and structural properties of the samples were investigated by scanning electron microscopy (SEM), x-ray diffractometry (XRD), and transmission electron microscopy (TEM). Plan-view scanning electron micrographs, covering several hundreds of NWs, were analyzed using the open-source software IMAGEJ [51] to assess the NW number densities and diameter distributions. In the case of ensembles with a low density of uncoalesced NWs, where the analysis of plan-view images is challenging due to NW tilt as well as the presence of a parasitic GaN layer, the diameter distributions were derived from the analysis of several cross-sectional scanning electron micrographs. Representative plan-view and cross-sectional scanning electron micrographs can be found in the Supporting Information. XRD experiments were performed with CuK$_{\alpha 1}$ radiation using a Panalytical X-Pert Pro MRD™ system equipped with a Ge(220) hybrid monocromator. Symmetric $\theta/2\theta$ scans and pole figures were recorded with a 1 mm slit in front of the detector. Cross-sectional TEM specimen were prepared using the standard method of mechanical grinding and dimpling down to below 25 µm followed by Ar-ion milling. Transmission electron micrographs were acquired using a JEOL JEM 3010 microscope operating at 200 kV and equipped with a Gatan charge coupled device camera. To assess the polarity of the samples, they were analyzed on micro- and macroscopic scales by convergent beam electron diffraction (CBED) and KOH chemical etching, respectively. CBED measurements were performed using a beam spot size of 6 nm and compared to simulations performed by JEMS.[52] To elucidate the polarity by chemical etching, the samples were exposed to a 5 M KOH aqueous solution at 40°C for 10 min. The interface between the Ga-polar GaN layer and the N-polar GaN NWs was investigated by high-resolution transmission electron microscopy (HRTEM) using a JEOL JEM 3010 equipped with a LaB$_6$ cathode and a UHR pole piece resulting in a point resolution of about 0.17 nm.

## Supporting Information

Effect of Si deposition on surface morphology; evolution of the reflection high-energy electron diffraction pattern during growth; additional cross-sectional transmission electron micrographs; and representative scanning electron micrographs used to assess the density and diameter of GaN NWs.

## Note on author contributions

Ziani de Souza Schiaber and Gabriele Calabrese contributed equally to this work.

## Acknowledgements

We thank Carsten Stemmler for his help to prepare the samples and his dedicated maintenance of the MBE system together with Hans-Peter Schönherr and Claudia Herrmann. We are indebted to Anne-Kathrin Bluhm for her technical assistance to acquire scanning electron micrographs, Johannes K. Zettler for his preliminary growth experiments, and Javier Bartolomé Vílchez for valuable discussions on HRTEM. Gabriele Calabrese acknowledges the financial support provided by the Leibniz-Gemeinschaft under Grant SAW-2013-PDI-2. We also thank Vladimir Kaganer for a critical reading of our manuscript.

**Table of Contents Graphic**

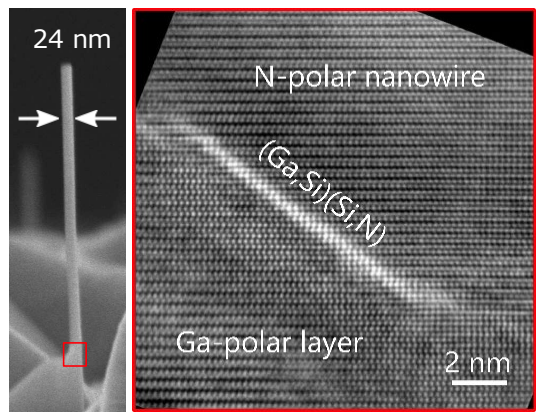